\documentclass[conference]{IEEEtran}
\IEEEoverridecommandlockouts
\usepackage{cite}
\usepackage{amsmath,amssymb,amsfonts}
\usepackage{algorithmic}
\usepackage{graphicx}
\usepackage{textcomp}
\usepackage{xcolor}
\usepackage{verbatim}
\usepackage{float}
\def\BibTeX{{\rm B\kern-.05em{\sc i\kern-.025em b}\kern-.08em
    T\kern-.1667em\lower.7ex\hbox{E}\kern-.125emX}}
\begin{document}

\title{Tomen: Application of Bitcoin Transaction Based on Tor}

\author{
\IEEEauthorblockN{1\textsuperscript{st} Yuanzhe Jin}
\IEEEauthorblockA{
\textit{Northwestern University}\\
Evanston, USA\\
yzjin@u.northwestern.edu}

\and
\IEEEauthorblockN{2\textsuperscript{nd} Ziheng Dong}
\IEEEauthorblockA{ 
\textit{Northwestern University}\\
Evanston, USA \\
zihengdong@u.northwestern.edu}

\and
\IEEEauthorblockN{2\textsuperscript{nd} Xing Li}
\IEEEauthorblockA{ 
\textit{Xiamen University}\\
Xiamen, China \\
lixing73@stu.xmu.edu.cn}
}
\maketitle

\begin{abstract}
Bitcoin has emerged in 2008, and after decades of development, it has become the largest trading currency by far. The core of the blockchain is to ensure the anonymity of user transactions. As more and more analysis algorithms for blockchain transactions appear, the anonymity of the blockchain is increasingly threatened.

We propose Tomen, an encryption application for the communication process in the bitcoin transaction process, combined with the encryption principle method of Tor. The goal is to achieve the application of the anonymization of bitcoin transaction communication.
\end{abstract}

 
\section{Introduction}
Tor is free software that enables anonymous communication. Its name is derived from "The Onion Router." Users can access Tor via free coverage provided by global volunteers to achieve the purpose of hiding the user's real address, avoiding network monitoring, and network traffic analysis. Tor users' Internet activities, including browsing online websites and instant messaging, are challenging to track. Tor was designed to protect users' privacy, as well as the freedom and ability to communicate without supervision secretly.

Tor does not prevent online websites from judging whether users access the site through Tor. Although it protects users' privacy, it does not hide the fact that users are using Tor. Some websites restrict the use of Tor. For example, Tor block restricts editing through Tor. 
 
Tor implements onion routing as a technology by encrypting at the application layer in the transport protocol stack. Tor encrypts the data, including the IP address of the next node multiple times, and submits it through virtual circuits (including randomly selected Tor nodes). Each relay decrypts a layer of encrypted data to know the next destination of the data, and then sends the remaining encrypted data to it. The final relay will decrypt the innermost encrypted data and send the original data to the destination address without leaking or knowing the source IP address.

Because of its anonymity, Tor is often used in accessing the deep web and dark web, and many people came to know Tor is from the news about Bitcoin. Some researchers suggested that the medical community must be aware that the development of the dark web and Bitcoin has made illegal activities easier. They discussed which healthcare fields are affected by these technological developments of the Internet and their consequences and expressed their views on what measures can be taken to protect the Internet community\cite{Masoni2016}.

Bitcoin has been wildly used in the dark web as a payment. Some malicious software creates a bitcoin wallet for each victim, and the payment method is in the form of digital bitcoin using Tor gateway\cite{7813706}. In some way, it can be considered that illegal transactions in the dark web have made Bitcoin known to more people, and thus made Bitcoin accepted by more people. Bitcoin is currently the most popular Cryptocurrency on the dark web, followed by Monero and Litecoin. According to calculations, on the dark web, all websites accept Bitcoin payments, and more than half of the sites provide exclusive Bitcoin support. Besides, the average number of cryptocurrencies supported by each market is approximately 2.4. And Monero is the only asset other than Bitcoin that also has the exclusive support of the dark web market.

Some studies provide a systematic explanation of the opportunities and limitations of anti-money laundering (AML) in Bitcoin. The study observed that Bitcoin attracted criminal activity. Although this statement has not undergone rigorous scrutiny, some services that enhance the anonymity of transactions have emerged in the Bitcoin ecosystem\cite{6805780}.
\section{Related Work}
Bitcoin has become the most successful cryptocurrency in history. Although only a rough analysis of the design of the system, the economic value of Bitcoin has increased by billions of dollars\cite{7163021}. In addition to attracting a billion-dollar economy, Bitcoin has transformed the digital currency space. In\cite{7423672}, Tschorsch et al. showed that the key ideas in Bitcoin are equally applicable to other fields, so their impact goes far beyond Bitcoin itself. Past research has shown that Bitcoin is not as anonymous as it seems, and the inherently public nature of blockchain technology makes it difficult to obtain privacy\cite{inproceedings}. Researchers have proposed solutions to the anonymity of transactions on and off the Bitcoin blockchain. Anonymous credentials can be issued by using an untrusted third party, and users can exchange them for bitcoin to achieve anonymity of transactions\cite{10.1007/978-3-662-53357-4_4}. Some other researchers find the measurement of various decentralized indicators of the two largest cryptocurrencies (Bitcoin and Ethereum) with the largest market value and user base. By measuring the network resources of the nodes and the interconnection between them, the protocol requirements that affect the operation of the nodes, and the robustness of the two systems against attacks are calculated\cite{10.1007/978-3-662-58387-6_24}.

Some papers show Tor can be attacked, which means it is not a safe way to protect the Bitcoin transaction. It was once considered subversive to be able to safely hide from network services\cite{jawaheri2018deanonymizing}. However, researchers have demonstrated a design flaw in Tor. In the research paper \cite{10.1007/978-3-642-14992-4_19}, they said that by analyzing the transmission of encrypted data by volunteers on the Tor network on a separate computer, attackers can infer the location of hidden servers or find designated Tor users through information sources. 
Other research\cite{7163022} also supports this idea, it proved that combining Tor and Bitcoins can create new attack vectors. A resource-poor attacker can fully control the flow of information between all users who choose to use Bitcoins instead of Tor. In this article\cite{b0421066bd0d4590953f8d336f99be4c}, it shows an attack on the transaction between Bitcoin and Tor itself. Besides Tor, researchers have proposed an effective method of de-anonymizing Bitcoin users, which allows users to link their pseudonyms to the IP address that generated the transaction\cite{10.1145/2660267.2660379}. Some researchers point out that the routing infrastructure itself can be attacked in the currency through the Internet\cite{7958588}. In response to Bitcoin's characteristics, Eyal et al. proposed an attack in which collusive miners gained more than their due share\cite{10.1007/978-3-662-45472-5_28}. This attack may have major consequences for Bitcoin: rational miners will be more willing to join selfish miners, and the size of the conspiracy group will continue to expand until it becomes a majority. At this point, the Bitcoin system is no longer a decentralized currency.

Researchers have discovered that a guard can easily find packet traffic passing in all directions\cite{10.1007/978-3-642-15497-3_16}. Using machine learning algorithms, it can be distinguished with high accuracy that this is an ordinary web page loop, and the introduction-point circuit is still a rendezvous-point. So there is no need to break Tor's encryption. Besides, by using Tor's computer to connect to a range of different hidden services, they showed that similar services to traffic analysis patterns can identify these services\cite{10.1007/978-3-030-02641-7_13}. This means that when a lucky attacker enters the position of the guard of the hidden service, it will be sure that it is the host of the hidden service\cite{10.1007/978-3-662-45472-5_30}. They can use bandwidth estimation methods to calculate the nodes in the Tor circuits and get the information from the users.

Above all, We can find that most researches pay attention to the attack on the encryption process. Our research aims to use Tor to build a useful application on information communication, which does not perform Tor encryption operations on the transactions themselves. 

\section{Tor Encryption Theory}
\subsection{Similarity and Difference between Blockchain and Tor}
First, the blockchain system and Tor are both distributed and fault-tolerant peer-to-peer networks. They use digital signatures to authenticate relevant participants, and they can update the system status by monitoring the network at any time to accept new transactions from nodes. The management of the network is mainly based on its architecture, and it also relies on cryptographic schemes for establishing Tor network circuits and verifying blockchain data. Also, the relevant peers retain their encryption keys to mutually authenticate and verify whether the information is accurate. Before performing key exchange on the Tor network, the distributed directory servers will first be contacted to obtain information about network members, but once the circuits are established, these servers will not interfere with communication\cite{8242004}.

Besides, blockchain technology is also used for zero-knowledge proof and information protection\cite{10.1257/jep.29.2.213}. The user can mathematically conclude that an input corresponds to a relevant definite output without revealing any additional information. Correspondingly, Tor’s goal is to protect users’ digital identities while disseminating user requests on public networks.

The main difference between the two is that the blockchain system maintains a secure global transaction ledger, while Tor is only used to forward user requests and has negligible visibility into network data. If Tor also has a consensus like a blockchain, then the key feature of the entire system-anonymity will disappear. At the same time, Tor is a layered network, where different nodes have different roles and different weights, and the maintenance of the system state is only reserved for some nodes that require user trust. Compared with the blockchain in design, the importance of general nodes is not the same. Tor only uses a limited number of nodes to control the network, so the flexibility of the Tor network is limited. An indispensable part is that Tor is managed by the Foundation in a less transparent manner, which means that the system security is built on a certain degree of unknown parts, especially the bridge-routing is currently completely unknown\cite{10.1007/978-3-642-39884-1_4}. The hidden service in the Tor network is a server that only accepts incoming connections through a specific protocol of the hidden service IP address to maintain its anonymity. Similarly, the bitcoin network can guarantee certain anonymity, because when forwarding transactions to multiple peers, there is no obvious connection between the node address and its IP.

Essentially, blockchain and Tor network provide decentralization, trust, resilience, and integrity. Besides, the difference is that Tor is not completely open like the blockchain so that any service can be built on the blockchain.

\subsection{Tor Network}
Tor is a triple proxy (when Tor sends a request, it will go through 3 nodes of the Tor network). There are two main server roles in its network:

One is the relay server. The relay server is a router responsible for transferring data packets, which can be understood as a proxy. The other is the directory server. It saves information about all the relay server lists in the Tor network (saves the relay server address and public key).

The Tor client first communicates with the directory server to obtain information about the globally active relay nodes, and then randomly selects three nodes to form a circuit. The user traffic jumps to these three nodes (hop) and finally reaches the target website server.

Tor's official website briefly introduces the principle of Tor. Tor is a triple proxy. The Tor client first communicates with the directory server to obtain global active relay node information, and then randomly selects three nodes to form a circuit. The user traffic jumps to these three nodes (hop) and finally reaches the target website. Server, so there are two entities in the Tor network, namely the user and the relay node. When users need to access the network anonymously, they first access the directory server to obtain information about Tor relay nodes worldwide, including IP addresses, public keys, egress policies, bandwidth, and online time. Then, three nodes are randomly selected to form a circuit, which are an ingress node, an intermediate node, and an egress node. When constructing the circuit, the user negotiates the shared session key with each relay node, and then sends the layer-encrypted information to the circuit. 

After each relay node decrypts it once, it sends the information to the next node. In this way, only the ingress node in the relay node knows the identity of the communication initiator. The intermediate nodes know the identities of the ingress and egress nodes in the channel, but they do not know the identities of the initiators and receivers of anonymous communication. The egress node is responsible for the application layer connection between the Tor network and the external Internet network, and acts as a relay between encrypted Tor network transmission traffic and non-encrypted Internet transmission traffic, and knows the identity of the anonymous communication receiver. Under this design, no node in the circuit knows the complete information, so anonymous communication is achieved.

\begin{figure}[htbp]
\centering
\centerline{\includegraphics[scale=0.30]{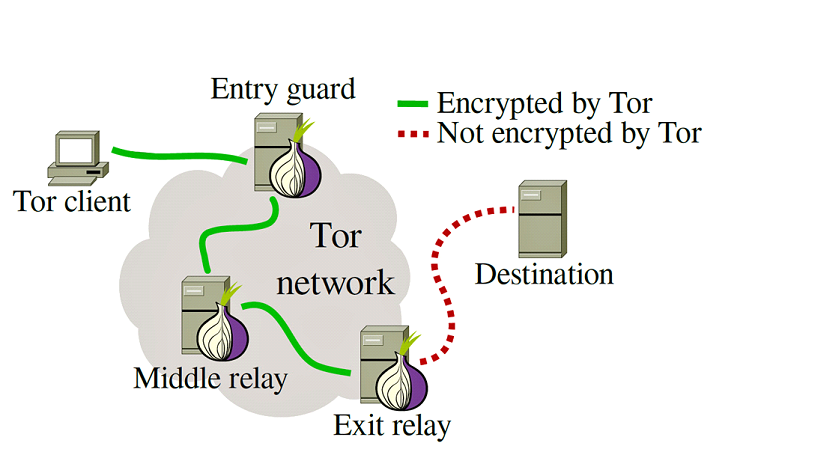}}
\caption{Tor protects the information with three nodes}
\label{tor network}
\end{figure}

Specifically, after the user starts Tor, the Tor client runs an Onion Proxy (OP) on the local machine and then starts to contact the directory server storing the global relay node information to obtain the global relay node information. After the OP obtains the relay node information, the OP will randomly select three nodes to form a circuit and negotiate the session key separately. In this process, each layer of the session is encrypted information once, until it is triple-encrypted, and only the egress node can see the plaintext. When the circuit is confirmed to be established, it starts to send real user access information. Besides, to enhance security, Tor reselects three nodes every ten minutes to avoid traffic analysis and honeypot nodes shown in the Fig.\ref{tor network}.

Fig.\ref{tor network} shows that the Tor network can encrypt between the client and its destination. Though it seems it does not encrypt some parts in the path, it is difficult and time-consuming for the attacker to use this non-encrypted network to get the essential information. As the attacker can only trace to the exit relay, it is always changing when the Tor sends the messages. 
\section{Deploy Tor in Blockchain Network}
In this section, we make some attempts to set up and run a Tor ‘relay’ with our hardware. According to its basic idea, the Tor network depends on volunteer users to gain more bandwidth. In other words, the more volunteers are using the network, the better it would be. The relays deployed by the users can help the Tor network to be faster, more robust against attack, and more stable.

According to the tutorial of the Tor community, there are three types of relays in the Tor network. All are no less important but surely they have different requirements in terms of hardware and legal implications. They are guard/middle relay, exit relay, and bridge relay, as mentioned in the previous sections. In our experiment, we choose to deploy a guard relay. The guard relay is the first relay in the Tor circuit of 3 relays. A middle relay, on the other hand, is not a guard or exit. It locates at the middle hop between the two.

Since we have a Tor network relay, we can develop some applications using the Tor network. The tool we utilize to help our development is ‘stem’. It is a Python package that enables users to interact with Tor. With stem, we can write scripts and applications that are similar to a typical server framework’s capabilities. In other words, the stem is an implementation of Tor’s directory and controllers using Python.

\begin{figure}[H]
\centerline{\includegraphics[scale=0.60]{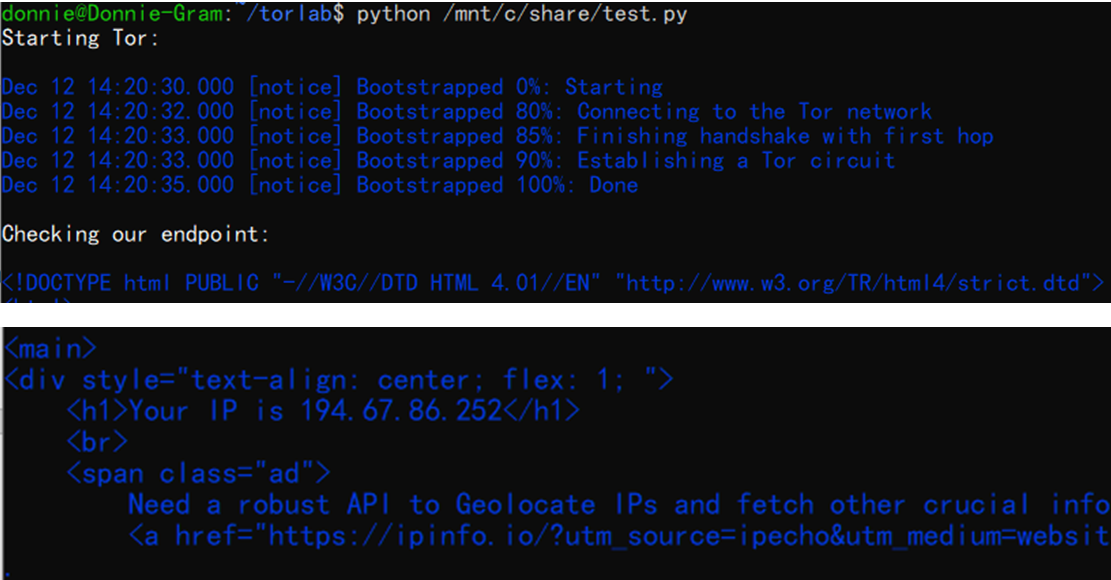}}
\caption{The echo shows that the Tor network works properly}
\label{result1}
\end{figure}

Firstly, we use this Python interface to start the Tor software on our server and check whether its functions are running appropriately. Here we build a query to a website that can echo the IP address of the user. The output goes as follows shown in Fig.\ref{result1}.

\begin{figure}[H]
\centerline{\includegraphics[scale=0.50]{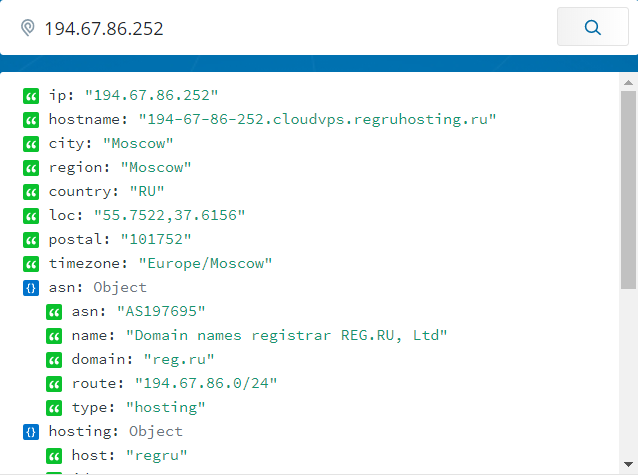}}
\caption{Hide IP to make the communication anonymous}
\label{result}
\end{figure}

Then we check the information about shown IP and find out our original IP information has been hidden behind the IP of another node. The information goes as follows in Fig.\ref{result}.

In Fig.\ref{result}, we can see that the IP address has been hidden from its origin, which helps the client to hide its location and also make the message annoymous.

\begin{figure}[H]
\centerline{\includegraphics[scale=0.28]{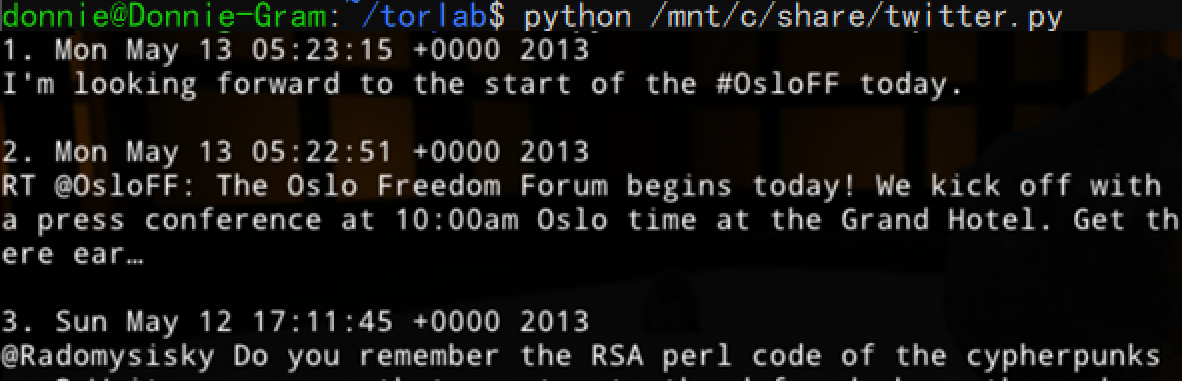}}
\caption{Use Twitter API to get encrypted communication between two clients}
\label{twitterapi}
\end{figure}

With these basic understandings, we can build an application to gain more information using the Tor network. Here we use the Twitter API as the example. Firstly, we need a register as a Twitter developer to use the APIs. Then we can generate tokens for the authentication. With these preparations done, we build the application on the same foundation as the previous one. We set up and start the Tor relay on our hardware. Then we set the configurations in the process of build the query. With these, we can call the API with the query we build and parse the response from it. The result is as follows shown in Fig.\ref{twitterapi}. 
 
As is shown in Fig.\ref{twitterapi}, we can get and parse the response from querying the API, with this capability we could build more secure and handful applications with the Tor network in the future.

\section{Conclusion}
The Tor network, with more than 2 million daily users, is the world’s most popular anonymous system for Internet users. It was once considered subversive to be able to safely hide from network services. 

Though there are many ways to de-anonymous Tor, it still should be considered to be the most secure way to anonymously access the Internet. With the explosive growth of Bitcoin in the dark web, Tor has gradually become popular among the general public. At the same time, Tor has also helped Bitcoin become a widely accepted payment method. But both have been criticized for their links to illegal activities. As Bitcoin gradually moves towards financial markets, it plays more of a financial instrument role than an illegal transaction payment method. 

In our research, we demonstrate Tomen, the application of Tor technology. The application can realize the encryption of the transaction communication process, and realize the anonymization of the communication address in the process of sending transaction information.

\nocite{6956580}
\nocite{10.1007/978-3-642-32946-3_29}
\bibliographystyle{IEEEtran}
\bibliography{ref}

\end{document}